\documentclass{ckm}                 

\confname{Workshop on the CKM Unitarity Triangle, IPPP Durham, April
  2003}

\usepackage{graphicx}
\usepackage{dcolumn}
\usepackage{bm}
\usepackage{epsfig}
\begin{document}

\def\st{\scriptstyle}
\def\sst{\scriptscriptstyle}
\def\mco{\multicolumn}
\def\epp{\epsilon^{\prime}}
\def\vep{\varepsilon}
\def\ra{\rightarrow}
\def\mee{M_{ee}}
\def\ppg{\pi^+\pi^-\gamma}
\def\pmm{\pi^+\mu^+\mu^-}
\def\kpmm{K^+\rightarrow\pi^+\mu^+\mu^-}
\def\kpme{K^+\rightarrow\pi^+\mu^+ e^-}
\def\pke3{K^+ \rightarrow \pi^0 e^+ \nu}
\def\ake3{K \rightarrow \pi  e \nu}
\def\ke3g{K^+\rightarrow\pi^0 e^+ \nu \gamma}
\def\kmu3{K^+\rightarrow\pi^0\mu^+ \nu}
\def\pee{\pi^+e^+e^-}
\def\kpee{K^+\rightarrow\pi^+e^+e^-}
\def\ke4{K^+\rightarrow\pi^+\pi^-e^+\nu}
\def\kpipi{K^+\ra \pi^+\pi^0}
\def\kmunu{K^+\ra \mu^+\nu}
\def\k3pi{K^+\ra \pi^+ \pi^0 \pi^0}
\def\ktau{K^+\ra \pi^+ \pi^+ \pi^-}
\def\eeg{e^+e^-\gamma}
\def\dal{\pi^0 \ra e^+ e^- \gamma}
\def\ddal{\pi^0 \ra e^+ e^- e^+ e^-}
\def\Mnu2{M_{\nu}^{2}}
\def\vp{{\bf p}}
\def\ko{K^0}
\def\kb{\bar{K^0}}

\title{
New, high statistics measurement of the $\pke3$ ($K^+_{e3}$) branching ratio
}

\author{
Julia A. Thompson,
D.E. Kraus, and 
 A. Sher\thanks{
Present address: SCIPP UC Santa Cruz, Santa Cruz, CA 95064.},
for the E865 collaboration}
\address{ Department of Physics and Astronomy, University of
Pittsburgh, Pittsburgh, PA 15260, USA }

\begin{abstract}
E865 at the Brookhaven National Laboratory AGS collected $\approx$ 70,000
$K^+_{e3}$
events to  measure the $K^+_{e3}$ branching ratio
relative to the  $\kpipi$, $\kmu3$, and $\k3pi$ decays. 
The $\pi^0$  was detected using the $e^+ e^-$ pair 
from $\dal$ decay and no photons were required.
Using the Particle Data Group branching ratios 
\cite{pdg} for the normalization decays we obtain 
$BR(K^{+}_{e3(\gamma)})=(5.13\pm0.02_{stat}\pm0.09_{sys}\pm0.04_{norm})\%$, 
where $K^{+}_{e3(\gamma)}$ includes the effect of virtual and real photons.
This result is $\approx 2.3\sigma$ higher than the current 
Particle Data Group value. 
Implications for the $V_{us}$ element of the CKM matrix,
 and the matrix's unitarity are discussed.
\end{abstract}


\maketitle


\begin{figure}
\hbox to\hsize{\hss
\includegraphics[width=\hsize]{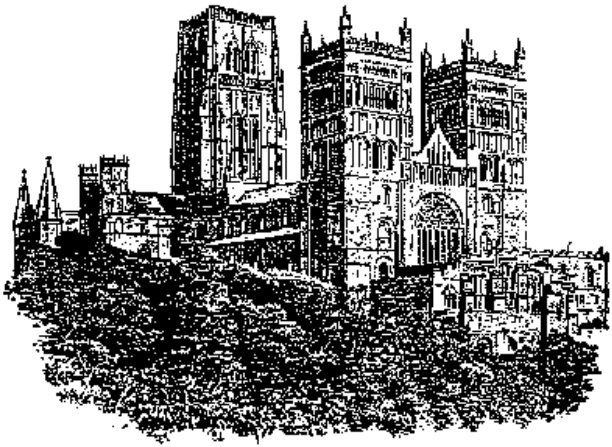}
\hss}
\caption{The cathedral at Durham.}
\label{fig:cathedral}
\end{figure}


   This contribution follows the thesis of Alexander Sher
\cite{ke3_th} and the E865 group preprint \cite{ke3_prl}
prepared  after the conference. I will comment on some
differences in presentation,
and emphasize  systematic checks.
Discussions during all the  conference activities (as for
example in Figure \ref{fig:cathedral}) were valuable in clarifying
the connection of our data to other experiments.

 The experimentally determined Cabibbo-Kobayashi-Maskawa (CKM) matrix
describes mixing between 
the "intrinsic" and physically observed quarks 
and is  unitary within the Standard Model.
One particularly interesting  unitarity condition involves the first row 
elements:
\begin{equation}
|V_{ud}|^2+|V_{us}|^2+|V_{ub}|^2=1-\delta
\label{unitest1}
\end{equation}
where  $\delta \ne 0$
 indicates a problem with unitarity, and 
 a sum less than one
 might  indicate  an
additional
quark generation.  Since LEP results seem to have ruled out more
than three neutrinos, such a shortfall  would  strain  the 
standard model.  Therefore, Eq. \ref{unitest1} 
has generated substantial interest.
The $V_{ud}$ element is obtained from nuclear and neutron decays.
While the precise shortfall in Eqn.  \ref{unitest1} varies with 
different determinations of $V_{ud}$, the qualitative
result is the same. For example, using an alternate,
precise, recent value 
from the nuclear superallowed Fermi beta decays leads to 
$\delta=(3.2\pm1.4)\cdot 10^{-3}$ \cite{hardy}.
$V_{ub}$, too  small to affect Eqn. \ref{unitest1}, is 
determined from 
 semileptonic decays of B mesons \cite{pdg}.

Although  $V_{us}$  can be determined either from hyperon or
from $\ake3$
decays, the purely vector $K_{e3}$ decays, with less  intrusion
of hadronic physics,  provide a
smaller  theoretical uncertainty\cite{pdg,lut}.
Theoretical contributions $V_{us}$  were reevaluated
recently\cite{rad,arie,calderon,bijnens}, but since 
uncertainties of $|V_{ud}|^2$ and $|V_{us}|^2$ are comparable,
a high statistics measurement of the $K^+_{e3}$ branching ratio
(B.R.) with good
control of systematic errors is welcome.

The bare (without QED corrections) 
$K^+_{e3}$ decay rate \cite{lut,rad,arie,dafne} can be expressed as
\begin{equation}
d\Gamma(K^{+}_{e3})={C(t)|V_{us}|^{2}}
|f_{+}(0)|^2[1+\lambda_{+}\frac{t}{M_{\pi}^2}]^{2}
dt
\label{ke3_rate}
\end{equation}
where 
$t=(P_{K}-P_{\pi})^2$,
C(t) is a known kinematic function, 
and $f_{+}(0)$ is the vector form factor value
at $t=0$ which has to be determined theoretically \cite{lut,rad}.
Two recent 
experiments\cite{kek,istra} provide $\lambda_+$ (the form factor slope)
measurements consistent with each other and with previous measurements.
An omitted negligible term in Eq. \ref{ke3_rate} containing the form
factor $f_{-}$ is proportional to $M_{e}^{2}/M_{\pi}^{2}$.

\begin{figure}[hbt]
\setlength \epsfysize{14cm}
\setlength \epsfxsize{7cm}
\includegraphics[scale=0.29]{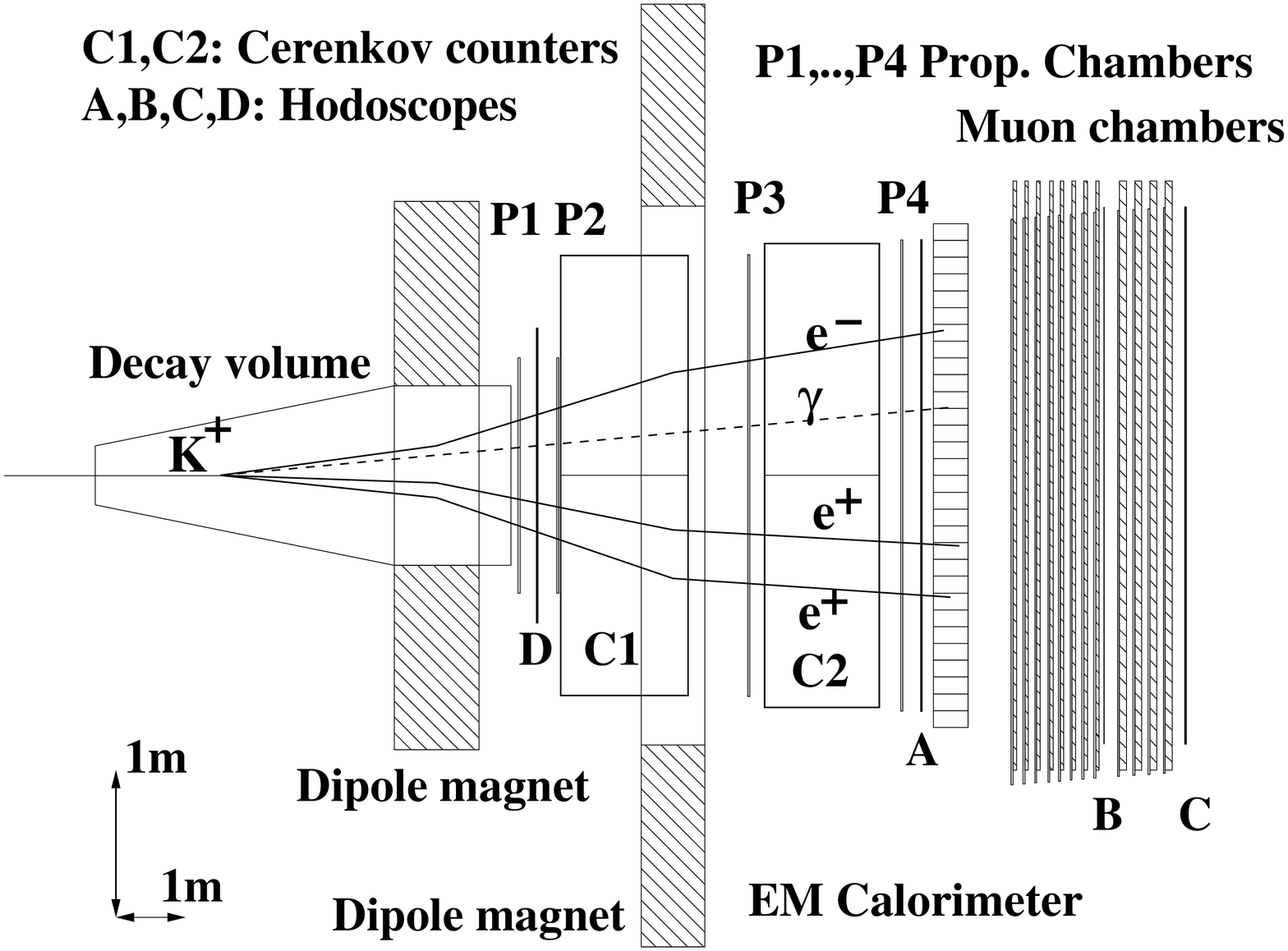}
\caption{Plan view of the E865 detector with a simulated $\pke3$ decay
followed by $\dal$.
}
\label{e865_det}
\end{figure}

E865 \cite{nim} searched for the lepton flavor number 
violating decay $\kpme$. The detector (Figure \ref{e865_det}) 
resided in a 6 GeV/c positive beam.
The first
dipole magnet separated decay products by charge. The second magnet
together with four multiwire proportional chambers (MWPCs: P1-P4)
formed the spectrometer. The particle identification system consisted of 
 threshold multichannel \v{C}erenkov counters (C1 and C2, each separated 
into left and right parts, for four independent volumes)
filled with gaseous methane ($\gamma_{t} \approx 30$, eff$_e \approx$
0.98\cite{ke3_th}),
an electromagnetic calorimeter, and a muon range system. The D and A 
scintillator
hodoscopes gave left/right and crude vertical position. The muon system
was not used for this analysis.

The decay  $\pi^0$ 
was detected through the $e^+ e^-$ pair from the $\dal$ decay mode, 
with the $\gamma$
detected in some cases.
In order to eliminate the uncertainty of the 
$\dal$ B.R. (2.7\%),
and to reduce experimental systematic uncertainty we used the other three 
major decay modes with a $\pi^0$
in the final state ($\kpipi$($K_{\pi2}$), $\kmu3$($K_{\mu3}$), 
$\k3pi$($K_{\pi3}$)) 
for the  normalization sample ("Kdal").

The $K^+_{e3}$ data were collected during a one-week dedicated run in
1998, with 
beam intensity reduced
to $\approx 10^7$ $K^+$/2.8 second pulse,
a factor of 10 less than the
standard $\kpme$ running intensity,
and  trigger logic prepared  especially  for this measurement.

The normalization  (Kdal) and  
$K^+_{e3}$ data  were collected by the "ELER"  trigger, which 
identified $e^+/e^-$ pairs (one  left,  one  right)
and required time coincidence of the four \v{C}erenkov counters with 
at least one 
D-counter scintillator slat on each (left and right) side of the detector.
The prescaled 
\v{C}erenkov efficiency trigger, requiring only three out of 
four \v{C}erenkov counters (no D-counter requirement),
was used to  measure the \v{C}erenkov and
D counter efficiencies. About 50 million triggers were accumulated,
with $\approx$  37 million  in the 
ELER trigger. The ELER events were 
about  3/4 accidentals, often with   a muon from high momentum beam 
particle decays $K\rightarrow \mu\nu$ or $\pi\rightarrow \mu\nu$
making  part of  the \v{C}erenkov portion of the trigger.

Off-line 
event reconstruction, using  the spectrometer
only, required a three charged track vertex inside the
decay
volume. 
The \v{C}erenkov and D counter efficiencies were obtained from the \v{C}erenkov
efficiency 
triggers. The redundancy of the MWPCs (4 planes/chamber), 
and   track reconstruction  was used to extract MWPC efficiencies.
The absence of the electromagnetic calorimeter from the trigger allowed 
its efficiency determination from the data.
Each  efficiency was measured over 
its appropriate  phase space.

The Monte Carlo simulation used GEANT \cite{geant}. All relevant kaon
decay modes were simulated, measured efficiencies were applied, and accidental
detector hits (from the data) were added to the simulated events.
The PDG value $\lambda_{+}=0.0278\pm0.0019$ \cite{pdg} was used for the
$K^{+}_{e3}$  simulation.

  The radiative corrections  are calculated both outside  
and inside the 3-body $K_{e3}$  Dalitz plot:
\begin{enumerate}

\item Outside: the $K^+_{e3\gamma}$ (inner bremsstrahlung and structure
dependent) decays which have photons hard enough to move the events
outside outside of the 3-body  
$K^+_{e3}$ Dalitz plot boundary. These  were explicitly simulated
\cite{dafne},
and are 0.5\% of the $K^+_{e3}$ process.
 This 0.5\%
is present in our data, increasing our observed events from the "bare"
$K^+_{e3}$,  but
 subtracted when
the "bare" $K^+_{e3}$ B.R. is calculated.

\item Inside: The difference  between the observed
events and the events from the bare $K^+_{e3}$ process.
Here we used the procedure of Ref. \cite{rad}. 
The observed events include four terms:
a) the bare process ($|a_0|^2)$; b) 
the $K^+_{e3\gamma}$ process ($|a_\gamma|^2$;  
c) the virtual photons $|a_v|^2$ correcting the bare process
(small compared to the other corrections), 
and d) the
interference of
the bare process  with the virtual photon corrections (2Re$a_0^*a_v$).  
  This difference is $-1.3\%$, i.e., the virtual photon
corrections and their interference with the bare process decrease
the number of physically observable events.  The overall difference
between the "observed, acceptance corrected" events and the events from
the "bare" process
is : 
$ N_{observed, ~acceptance~ corrected} =  N_{bare} (1. + 0.005 - 0.013)$.
 
\end{enumerate}

An overall correction of 1.0232 (the short distance
enhancement) is also applied\cite{sirnew,arie,rad}. Following
Cirigliano, et al.\cite{rad}, we  do not
apply this to our data in calculating the B.R.,  but apply it
in the calculation of $V_{us}$ from the B.R..

  The difference, inside the Dalitz plot,  between our detector's
acceptance for the
"bare" Ke3 process and for the observed process
 is an increase of  0.5\%.
Initially we expected the acceptance to decrease, since radiative
corrections shift data to lower electron energies, where we lose events.
However, we also have decreased acceptance at  high  electron
energies, where the electron energy distribution itself peaks.

For the $\dal$ decay, radiative corrections have been taken into account
according to Ref. \cite{mik}.

Table \ref{tab:dataflow} shows how the data were reduced
by event selection.
\begin{table}[hbt]
\caption{Effect of data selection cuts on sample size (1998 data).
 There is some overlap (which is accounted for in 
the analysis)  in the final
Ke3 and
Kdal  sample because of the third track  selection criteria.
 The 
 $K\rightarrow \pi^+\pi^0$ acceptance
 is $\approx 1.2\%$.  
Ke3 acceptance is  $\approx$ 0.7$\%$, somewhat lower because of the lower
average $e^+$
momentum in the Ke3 decay.  The acceptance can be approximately 
understood by taking a factor of three loss for each charged particle,
30$\%$ for the \v{C} ambiguity cut, and approximately a factor
of 2 for other quality cuts }
\begin{tabular}{|l|p{50pt}|}
\hline
 Selection Criterion &
Remaining Sample Size\\
\hline
Online trigger (2 tracks) & 37,676K \\
\hline

 2 track vertex    & $\approx$9000K\\
\hline

 3 track vertex			  & 2852K\\
\hline

 \v{C}Ambiguity 		  & $\approx$ 1900K \\
\hline
Vertex, aperture cuts             & 710 K \\
\hline
  $M_{ee} < $ 0.05 GeV/c$^2$       & 644K \\
\hline

 Low mass e+e- particle ID	& 626K \\
\hline
\hline
 Particle ID of third track (e/$\pi$,/$\mu$)       \\
\hline

   "Normaliz."(Kdal):$\pi\pi$,Kmu3,$\pi\pi\pi$   &  Ke3   \\

\hline

   $\approx$ 555K &  71K \\

\hline
\end{tabular}
\label{tab:dataflow}
\end{table}
Selection criteria, common to $K^+_{e3}$ and Kdal samples, included 
requirements for a good quality vertex, for the
three tracks to cross the active parts of the detector,  for the low
($M_{ee}<0.05$ GeV) mass $e^+e^-$ pair to be identified in 
the \v{C}erenkov counters, and for the second positive track to have less
than 3 GeV/c momentum.  The momentum cut avoids any situation in which 
a mode other than Ke3 can have the second positive track from a normalizer
mode properly satisfy the Ke3 criteria.  
A
geometric \v{C}erenkov ambiguity cut
rejected events where the \v{C}erenkov counter response could not be 
unambiguously
assigned to separate tracks ($\approx$ 30$\%$ loss of both $K_{e3}$ and
Kdal).

The $K^+_{e3}$ sample was then selected by requiring the second positive 
track to 
be identified as $e^+$ in 2 of the 3 electron detectors: C1, C2,
and the  calorimeter. 
Events entering the normalization sample
had no response in at least one of the two \v{C}erenkov counters.
These criteria minimized systematic uncertainties \cite{ke3_th}, but resulted
in a small overlap of the  $K^+_{e3}$ and Kdal samples, which 
was taken into account in the final result calculation.
Final acceptances for the three charged particles 
differed by no more than 4\% among  the three normalization
decays\cite{ke3_th}.

The final signal and normalization samples were
71,204 and 558,186, respectively.
Extensive comparisons between quantities in the data and simulation
 were presented in the thesis \cite{ke3_th} 
and the HEP preprint \cite{ke3_prl}.


Contamination of the $K^+_{e3}$ sample by other $K^+$ decays
occurred when  $\pi^+$ or
$\mu^+$
from normalization decays were misidentified as $e^+$ from $K^+_{e3}$, or
as a result of $\pi^0\rightarrow e^+ e^- e^+ e^-$ decay.
Care was taken with the 
PWC simulation so that track chisquares and vertex distributions agree
between data and Monte Carlo for well-measured tracks.  Decays ($\approx$
10$\%$ of $\pi$'s) are modelled in the simulation. Agreement of the 
standard vertex quality between data and simulation indicates successful
modelling of the decays at the level required for our measurement.
Systematic uncertainties were estimated by variation of the vertex quality
cut.

\begin{figure}
\setlength \epsfysize{8cm}
\setlength \epsfxsize{10cm}
\includegraphics[scale=0.30]{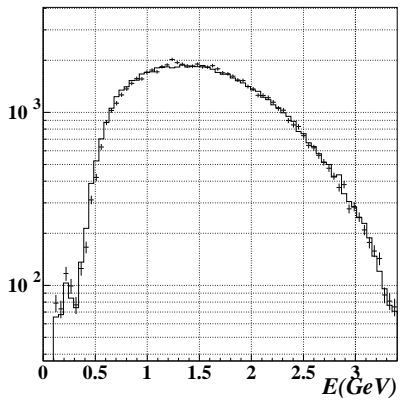}
\caption{
Energy deposited in the calorimeter by the second positive track from 
the selected $K^+_{e3}$ sample ($e^+$ which is not from the
low mass $e^+e^-$ pair).
No calorimeter information was used for the $e^+$ identification.
Markers with errors represent data; the histogram is simulation.
}
\label{e3_ke3_22}
\end{figure}

Total contamination of the Ke3 sample  was estimated to
be $(2.49\pm0.05_{stat}\pm0.32_{sys})\%$, with 
the systematic uncertainty caused by the simulation accuracy of 
the \v{C}erenkov counters' response to $\pi^+$ and $\mu^+$.
Contamination due to overlapping events was $(0.25\pm0.07)\%$ and 
$(0.12\pm0.05)\%$ of the selected normalization and 
signal samples respectively.
Figure \ref{e3_ke3_22} shows energy deposited in the calorimeter by the
$e^+$ from the selected $K^+_{e3}$ sample. The contamination
is manifest in the minimum ionization spike at 250 MeV.
The small excess of  data
in the spike
agrees with our contamination uncertainty estimate.

The final $K^+_{e3}$ sample included $\approx$30\% events with 
the fully reconstructed $\pi^0$. We used these events
as a consistency check but did not require photon detection in
our main  analysis.
This eliminated an additional systematic uncertainty from
photon detection and reconstruction in the calorimeter. However,
lack of a $\pi^0$ reconstruction increased vulnerability
to contamination from upstream decays and photon conversion.
Upstream decays whose photon produced pairs
before the decay volume were suppressed by requiring
the three track vertex to be more than two meters downstream of
the decay volume entrance. In addition, 
the results obtained from the two independent
samples (one with  and one without the $\pi^0$ reconstructed)
did not show a statistically significant discrepancy.
The decay volume was evacuated to about 
$10^{-8}$ nuclear interaction length, which suppressed
beam pion interactions.

The $K^+_{e3}$ sample size 
gives  a statistical precision of $0.4\%$.
The systematic error was determined from  stability of the result under
variation of the reconstruction procedure, 
selection criteria, detector efficiencies 
applied to the Monte Carlo, and subdivision (in possibly biased
distributions)  of 
both signal and normalization samples . No significant correlations
between 
different systematic uncertainties were observed.
Systematic uncertainties are summarized in
Table \ref{tab:sys}.

\begin{table}[hbt]
\caption{Systematic uncertainty sources and estimates of their respective
contributions to the error of the final result. The total error
was calculated as 
the sum of errors taken in quadrature.
}
\begin{tabular}{|l|p{50pt}|}
\hline
 Source of systematic error & Error estimate\\
\hline
 Magnetic field           & 0.3$\%$ \\
\hline

 Vertex, quality     & 0.6$\%$\\
\hline

 Vertex position			  & $0.2\%$\\
\hline

 \v{C}erenkov Ambig.		  & $0.3\%$ \\
\hline

  $M_{ee}$ cut                      & $0.2\%$ \\
\hline

  Aperture		  & $0.2\%$ \\
\hline

 $(\pi/\mu)^+$ iden.      & $0.04\%$ \\
\hline

 MWPC effic.		  & $0.2\%$ \\
\hline

 D ctr. effic.		  & $ 0.15\%$\\
\hline

 \v{C}erenkov effic.		  & $0.3\%$ \\
\hline

 Sample contam.
                                & $0.3\%$  \\
\hline

 Vertical  distrib. &  
                                   $0.8\%$  \\
\hline
    
 $e^+/e^-$ momentum distrib.
                          	  & $1.3\%$  \\
\hline

 $K^{+}_{e3}$ trigger effic.
                              	  & $0.1\%$  \\
\hline

  $K^+_{e3}$ f. f. ($\lambda^+$)
                              	  & $0.1\%$  \\
\hline

\hline

Total uncer.		           &  $1.8\%$ \\
\hline
\end{tabular}
\label{tab:sys}
\end{table}

\begin{figure}[hbt]
\begin{center}
\includegraphics*[scale=0.415]{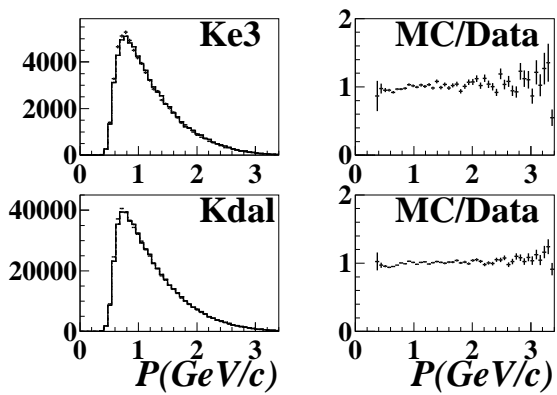}
\end{center}
\caption{Reconstructed momentum of the $e^+$ from
the low mass $e^+ e^-$ pair for the selected $K^+_{e3}$ and 
Kdal samples.
Histograms represent Monte Carlo;
points with errors are data. 
Monte Carlo to data ratios are shown on the right.}
\label{p12_ke3}
\end{figure}

The two largest contributions to the error come from the discrepancies 
\cite{ke3_th} between data and Monte Carlo in the momentum 
(Figure \ref{p12_ke3}) and spatial distributions \cite{ke3_th}.
The systematic 
error was determined by dividing $K^+_{e3}$ and Kdal
events in roughly equal samples using the relevant parameter as a
separator
and observing the result variation\cite{ke3_th}.
  The sensitivity of the vertical spatial discrepancy to 
the MWPC alignment and of the momentum discrepancy to the spectrometer 
parameters
\cite{ke3_th} indicate possible origins of these discrepancies.

In addition to the  check using  fully reconstructed $\pi^0$'s, as  a 
second 
consistency check, we estimated the $\ktau$ decay 
B.R. relative to the Kdal  sample.
 The result
was $(1.01\pm 0.02)\times R_{PDG}$, where $R_{PDG}$ is the prediction based
on the PDG compilation, and the theoretical prediction \cite{dalitz} was used 
for the $\dal$ decay rate. The 2\% error  was determined by combining
all relevant uncertainties in quadrature, and was dominated by the uncertainty
in the prescale factor of the trigger used to collect $\ktau$ events.
This result checks both the treatment of decays and the particle 
identification.

A third  check compared the $K^+_{e3}$ B.R.
from 1998 and  1997   data.
The 1997 $K^+_{e3}$ data used a trigger  requiring
hits in the calorimeter, A and D-counters. That trigger  neither
allowed  measurement  of these detector efficiencies,
nor  of the $K^+_{e3}$ trigger 
efficiency.
While we did not use the 1997 data for our final result, 
the 1997 $K^+_{e3}$ B.R. was statistically consistent
 (within one sigma) with
that  from the 1998 data.
This agreement is  important since the momentum
comparison in the 1997 data 
looks qualitatively 
different from the 1998 data\cite{ke3_th}. 
A preliminary
 reconstruction version was used for the 1997 data,
without the final magnetic field tuning and detector
realignment. Our
intuition is that the  discrepancies in 
decay product momenta  and spatial distributions,
which dominate the systematic uncertainties, reflect
 our imperfect knowledge of the magnetic field and
detector positions but do not bias our result beyond our
estimated systematic errors.  Further
refining of this knowledge was
judged to be a Herculean task (the magnet, e.g., had
been decommissioned) and was not undertaken.

We estimated the form factor slope $\lambda_+$ from both 1998 and 1997
$K^+_{e3}$ data samples. We obtained: $\lambda_{+}=0.0324\pm0.0044_{stat}$ for the 1998,
and $\lambda_{+}=0.0290\pm0.0044_{stat}$ for the 1997 data, both 
consistent with the current PDG fit.
%

After subtraction of  contamination in the Ke3 sample \cite{ke3_th},
our result is $BR(K^{+}_{e3(\gamma)})/(BR(K_{\pi 2})+BR(K_{\mu 3})+
BR(K_{\pi 3}))
=0.2002\pm0.0008_{stat}\pm0.0036_{sys}$,
where $K^+_{e3(\gamma)}$ 
includes all QED contributions (loops and inner bremsstrahlung).
As noted above, for this result the $\pi^0$ was detected using 
the $e^+ e^-$ pair from $\dal$ decay and no photons were required.

Using the PDG fit \cite{pdg} values for the normalization branching ratios
we infer 
$BR(K^{+}_{e3(\gamma)})=(5.13\pm0.02_{stat}\pm0.09_{sys}\pm0.04_{norm})\%$ 
where the normalization error is determined by the PDG estimate
of the normalization  B.R. uncertainties.
The PDG fit to the results to the previous $K^+$ decay 
experiments yields $BR(\pke3)=(4.87\pm0.06)\%$ \cite{pdg},
$\approx 2.3\sigma$ lower than our result.

As discussed above,  the total radiative correction was 
$0.8\%$, yielding 
$BR(\pke3_{bare})=(5.17\pm0.02_{stat}\pm0.09_{sys}\pm0.04_{norm})\%$.
This  differs slightly from the thesis  (5.16\%)
due to recalculation with the 2002 PDG, while the thesis used  the 2001 
PDG values.

Using the current PDG value for $G_{F}$, 
the short-distance enhancement factor 
$S_{EW}(M_{\rho},M_{Z})=1.0232$\cite{rad,sirnew},
and our result for the bare $K^+_{e3}$ decay rate we obtain 
$|V_{us}f_{+}(0)|=0.2239 \pm 0.0022_{rate} \pm 0.0007_{\lambda_{+}}$, 
which leads to 
$|V_{us}|=0.2272 \pm0.0023_{rate} \pm 0.0007_{\lambda_{+}} \pm 0.0018_{f_{+}(0)}$
if $f_{+}(0)=0.9874\pm0.0084$\cite{lut,rad}. With this value of $V_{us}$
and $V_{ud}$ from superallowed nuclear Fermi beta decays\cite{hardy}, we
obtain
$\delta=0.0001 \pm 0.0016$.

This result is consistent with  CKM 
unitarity, but our $K^+_{e3\gamma}$ result is $5.3\%$  higher than the PDG
2002 value. with a
statistical
error $0.4\%$, systematic error $ 1.8\%$, and normalization 
error $0.7\%$. 

We conclude:
a) the Ke3 B.R. may be somewhat higher
than that listed in the PDG tables; and  
b) Ke3 experiments are now precise enough that  radiative corrections
should be applied consistently  to all entries
 in the PDG average.

 Even without our result, the V$_{us}$ extracted from the PDG average
$K^+_{e3}$ decay rate
is  higher than  the $V_{us}$ from the  $K^0_{e3}$ rate.
\cite{Cirigliano,kloe}.
Our $K^{+}_{e3}$  result increases this difference. 
$K_{e3}$ decay measurements (both charged and neutral) in progress 
(CMD2, NA48, KLOE)\cite{kloe}
should  clarify the experimental situation.

We thank V. Cirigliano for the $K^+_{e3}$ radiative corrections
code and insights concerning the physical source of the radiative
corections. We also thank,
 A. Baratt, G. Isidori, P. Lichard,  S. Eidelman,
A. Poblaguev, and 
B. Schwartz for vigorous and helpful discussions.
We gratefully acknowledge all our E865 colleagues /cite{nim} and the 
contributions by the  
 staff  of the AGS 
and 
 participating institutions, particularly Bob Giles, and
students Tim Stever,
Cindy Miller, Beth Scholle, Tuan Lu, Elisabeth Battiste, John
Hotmer, and Melinda Nickelson,  among the many
who helped to make  the Cerenkov counters, and to make them 
effective. This
work was supported in part by 
the U.S. Department of Energy, the National Science Foundations of 
the USA, Russia and Switzerland, and the Research Corporation.

\end{document}